# Boron-doped graphene – DFT study of the role of dopant concentration and oxidation on sodium and aluminium storage applications

M.S. Ritopečki[1], N.V. Skorodumova[2,3], A.S. Dobrota[1*], I.A. Pašti[1]

[1] *University of Belgrade – Faculty of Physical Chemistry, Belgrade, Serbia*

[2] *Department of Materials Science and Engineering, School of Industrial Engineering and Management, KTH - Royal Institute of Technology, Stockholm, Sweden*

[3] *Applied Physics, Division of Materials Science, Department of Engineering Sciences and Mathematics, Luleå University of Technology, Sweden*

**Abstract**

Graphene is thought to be a promising materials for many applications. However, pristine graphene is not suitable for most electrochemical devices, where defect engineering is crucial for its performance. We demonstrate how boron doping of graphene can alter its reactivity, electrical conductivity and potential application for sodium and aluminium storage, with the emphasis on novel metal-ion batteries. Using DFT calculations, we investigate both the influence of boron concentration and the oxidation of the material, on the mentioned properties. It is demonstrated that the presence of boron in graphene increases its reactivity towards atomic hydrogen and oxygen-containing species, in other words, it makes B-doped graphene more prone to oxidation. Additionally, the presence of these surface functional groups significantly alters the type and strength of the interaction of Na and Al with the given materials. Boron-doping and oxidation of graphene is found to increase Na storage capacity of graphene by the factor of up to 4.

---------------------------------------------------------------

* Corresponding author email: ana.dobrota@ffh.bg.ac.rs

# 1. Introduction

Graphene is a two-dimensional material with a graphite monolayer structure and $sp^2$-hybridized carbon (C) atoms arranged in a honeycomb-like lattice [1,2]. Due to its characteristics, such as good electric conductivity and a large surface area, graphene has attracted interest as a potential anode material. However, the fact that it's a zero-bandgap semiconductor, along with its chemical inertness, limits its application possibilities [3–7]. Therefore, the graphene lattice is usually changed by chemical doping or by introducing covalent bonds with certain chemical groups or even molecules [8,9]. It should, however, be noted that the defects in graphene can also occur spontaneously during synthesis, and they can be sometimes difficult to predict and control [10]. The simplest modification method of pristine is chemical doping by heteroatoms, which improves the properties of pristine graphene by improving the metal-surface interaction, ~~and~~ adsorption and charge transfer abilities [2,4,11,12]. Heteroatoms such as boron (B), nitrogen (N), sulfur (S), fluorine (F), and phosphorus (P) are frequently chosen for this purpose since they change the bandgap and the position of the Fermi level, which depends on their concentration [4,13].

Carbon-based materials, including graphene, are widely investigated as candidates for electrode materials in metal-ion batteries (MIBs). For MIBs, the strength and the nature of interactions of the metal (and/or metal-based ions) with electrode materials are essential. Novel types of MIBs focus on metals other than Li as the electroactive species, due to the problems associated with Li-ion batteries [14] and Li abundance in general. These other metals include Na and Al, as they have small mass and radius, and high abundance in Earth's crust. Additionally, further advantages of Al are its three-electron redox property and low flammability. However, both Na and Al interact with pristine graphene relatively weakly [15], resulting in low voltages of hypothetical novel MIBs with graphene as the electrode material. Thus, chemical functionalization is needed to boost graphene's Na and Al storage capacity. Some reports of Al-ion batteries, which use carbon-based electrodes can already be found in the literature [16,17].

Boron-doped graphene is an especially good candidate for doping of graphene since B- and C-atoms have similar radii, making a C−B bond in a monolayer only ~0.5 % longer than the C−C one. Because of this, doping does not affect the planarity of the graphene sheet or the $sp^2$ hybridization of the carbon atoms within [11,18]. Boron is less electronegative than carbon, which causes a difference in electron density between boron and carbon atoms, leaving the boron site with an electron deficit [19]. In addition, since boron has one electron less, B-graphene acts as a p-type semiconductor [20]. B-graphene also has, similarly to pristine graphene, a very low or non-detectable total magnetization [12,20]. For now, B-graphene has primarily found its place in energy-related applications, such as batteries and supercapacitors [21,22], but there are also studies where it is used as a gas and optical sensor material [23,24] and in the biomedical field of study [25]. For example, Ling and Mizuno used first-principles calculations to demonstrate that the sodiation of boron-doped graphene preserves its structural integrity, avoids formation of dendrites and allows for approx. 2 times the capacity of the graphite anode in Li-IB and approx. 2.5 times higher than hard carbon in Na-IB [26]. Recently, Yiqun *et al.* demonstrated that the cathode based on boron-doped reduced graphene oxide displays a high Al-storage capacity and outstanding long-term stability [27].

In this study, we continue our previous work [28], where it was shown that the introduction of dopants into oxidized graphene, or controlled oxidation of doped graphene can enhance Na storage capabilities of the material, with boron found to be a very suitable dopant for this purpose. Thus, we investigate the effects of different concentrations of boron as a dopant in graphene on its properties, which are of importance for electrochemical applications. First, we focus on the reactivity of the material towards H, O and OH, and next we investigate the adsorption of Na and Al onto bare and oxidized boron-doped graphene, to explore the possibility of using such modified graphene surfaces in novel energy storage devices.

## 2. Computational details

In order to carry out the DFT calculations, open-source program packet Quantum ESSPRESO [29,30] was used. GGA and PBE functionals were used as part of this package. Since the adsorption of sodium and aluminium on graphene-based materials was investigated, dispersion interactions need to be included by utilizing DFT-D2 correction [31]. The kinetic energy cut-off of plane waves was set at 36 Ry and the density cut-off was set at 576 Ry. Spin polarization was included in all the conducted calculations. The first irreducible Brillouin zone was obtained using a Γ-centered 4×4×1 grid utilizing the general Monkhorst–Pack scheme [32]. Löwdin charges were used for discussing charge transfer. To investigate electronic structure a denser centred 20×20×1 grid was used.

Pristine graphene was modelled as a $(3\sqrt{3}\times3\sqrt{3})R30°$ structure containing 54 carbon atoms, $C_{54}$. Boron-doped graphene was constructed by replacing one, two, or three carbon atoms with boron, obtaining graphene with 1.85, 3.70, or 5.56 at.% of B, respectively. To evaluate the stability of boron-doped graphene, the binding energy ($E_b$) of boron at the vacancy sites is used. The binding energy of the $n^{th}$ ($n\in\{1,2,3\}$) substitutionally introduced boron atom is calculated as:

$$E_b(n^{th}\,B) = E(C_{54-n}B_n) - E(C_{54-n}B_{n-1}) - E(B) \quad (1)$$

where $E(C_{54-n}B_n)$ stands for the total energy of graphene doped with $n$ atoms of B, $E(C_{54-n}B_{n-1})$ the total energy of graphene doped with $n-1$ atoms of B and a vacancy at the site where the $n^{th}$ B will be introduced, and $E(B)$ the total energy of an isolated boron atom. Since the adsorption of atomic hydrogen, atomic oxygen, hydroxyl group on $C_{54-n}B_n$ was also investigated, their corresponding adsorption energies ($E_{ads}$) were calculated using the following formula:

$$E_{ads}(A) = E_{subs+A} - E_{subs} - E_A \quad (2)$$

where $E_{subs+A}$, $E_{subs}$, and $E_A$ stand for total energies of the optimized adsorbate@substrate system, bare substrate ($C_{54-n}B_n$), and isolated adsorbate atom/group (A = H, O, or OH), respectively. The most stable O@$C_{54-n}B_n$ and OH@$C_{54-n}B_n$ structures were further used as models of oxidized $C_{54-n}B_n$.

Finally, adsorption of Na and Al on bare $C_{54-n}B_n$ and its oxidized forms (O@$C_{54-n}B_n$ and OH@$C_{54-n}B_n$) were investigated, while the strength of their interaction with the substrate was estimated in terms of differential and integral adsorption energies. The differential adsorption energy ($E_{ads,diff}$) was calculated as:

$$E_{ads,diff}(M) = E_{subs+mM} - E_{subs+(m-1)M} - E_M \quad (3)$$

where "subs" can be $C_{54-n}B_n$, O@$C_{54-n}B_n$ or OH@$C_{54-n}B_n$, while $E_{subs+mM}$ stands for the total energy of the chosen substrate with $m$ atoms of metal M adsorbed (i.e. $m$ is the number of atoms M, M = Na or Al). Similarly, $E_{subs+(m-1)M}$ stands for the total energy of the chosen substrate with $m-1$ atoms of metal M adsorbed. The integral adsorption energy ($E_{ads,int}$) was calculated as:

$$E_{ads,int}(M) = (E_{subs+mM} - E_{subs} - m\cdot E_M) / m \quad (4)$$

where $E_{subs}$ stands for the total energy of the optimized substrate without any M adsorbed. Charge redistribution caused by metal interaction with the chosen model systems was investigated through a 3D plot of charge difference ($\Delta\rho$), defined as:

$$\Delta\rho = \rho_{\text{subs+M}} - \rho_{\text{subs,frozen}} - \rho_{\text{M}} \qquad (5)$$

where $\rho_{\text{subs+M}}$, $\rho_{\text{subs,frozen}}$ and $\rho_{\text{M}}$ stand for the ground state charge densities of the substrate interacting with M, the ground state charge densities of the substrate when M is removed (with frozen geometry), and that of the isolated M atom, respectively.

Graphical representations of all the graphene structures in this paper were made using VESTA [33].

## 3. Results and discussion

### 3.1. Graphene doping by boron

Pristine graphene, $C_{54}$, was used as a starting model. Its optimization resulted in C−C bond length of 1.43 Å, in agreement with previous studies [12,34,35]. B-doped graphene was modelled by replacing $n$ carbon atoms by boron, as described in the Computational Details section. While the position of the first boron atom can be chosen arbitrarily, it is necessary to perform additional calculations to find the optimal positions of the second and the third boron atom relative to the first one, or the first and the second one, respectively. In the case of $C_{52}B_2$ systems, we find that energetically least stable configuration is the one where the two B atoms are next to each other (**Fig. 1**). The systems become more stable as the two B atoms separate, and the optimal structure corresponds to the state where they are one across of the other, on the opposite sides of the same $C_4B_2$ hexagon (**Fig. 1**).

A similar pattern was found upon the introduction of the third boron atom – the dopant atoms are one across of the other, on the opposite sides of one hexagon, while it is optimal to have two boron atoms inside one hexagon. The first boron atom (the case of $C_{53}B$) is found to incorporate into the graphene lattice with the binding energy of –12.90 eV and C−B bond length of 1.49 Å, in agreement with our previous studies [12,36]. When more B atoms are introduced, the binding energies follow the trend: $E_b(1^{st} B) < E_b(2^{nd} B) < E_b(3^{rd} B)$, *i.e.* the first B is bound the strongest. The differences in their $E_b$ are up to 6.2%. The results are summarized in **Table 1**. All the investigated $C_{54-n}B_n$ systems were found to be nonmagnetic. We observe that there is no disruption of the planarity, even with the addition of three B atoms, which is a consequence of similar atomic radii of B and C. Upon the introduction of boron, the Bader charge of its first neighbouring carbon atoms increased by 0.1 e. On the other hand, the charge of boron atoms decreased by 0.1-0.2 e, which is expected since boron is less electronegative than carbon.

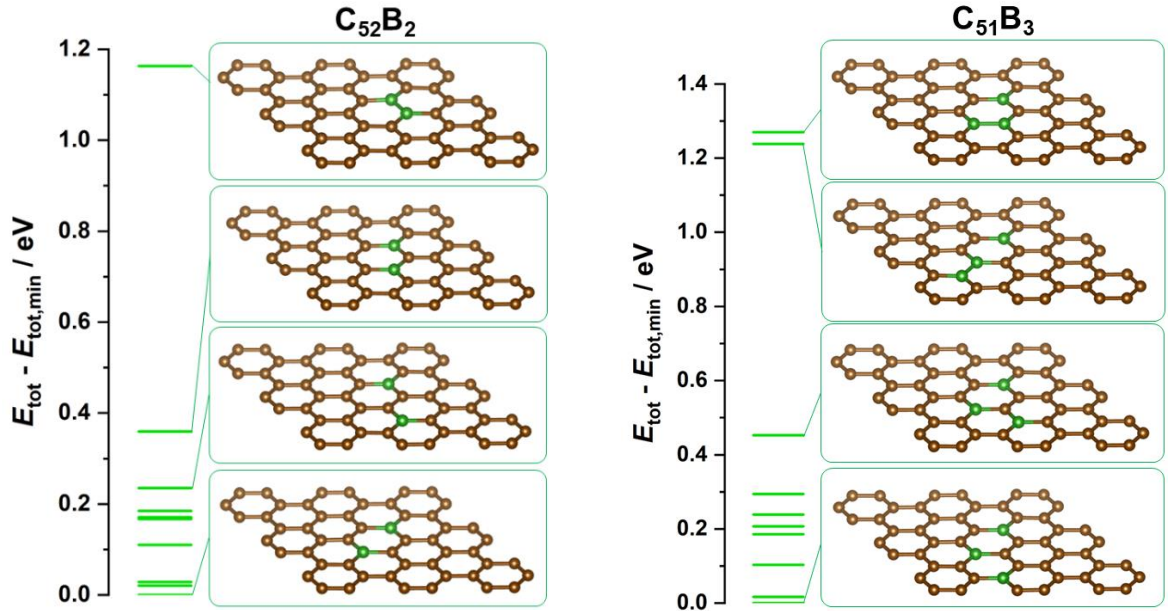

**Fig. 1.** Relative stabilities (in terms of total energies, $E_{tot}$) of the investigated $C_{52}B_2$ (left) and $C_{51}B_3$ (right) structures, compared to the most stable one, with $E_{tot,min}$, which corresponds to $E_{tot} - E_{tot,min} = 0$.

**Table 1.** Binding energies of boron atom ($E_b(B)$), C−B bond lengths ($d$(C−B)), and partial Löwdin charges of B ($\Delta q(B)$) for the studied $C_{54-n}B_n$ models.

| Model | $E_b(B)$ / eV | $d$(C−B) / Å | $\Delta q(B)$ / e |
|---|---|---|---|
| $C_{53}B$ | −12.90 | 1.49 | −0.1 |
| $C_{52}B_2$ | −12.46 | 1.51 | −0.2 |
| $C_{51}B_3$ | −12.15 | 1.51 | −0.2 |

Upon boron insertion into the graphene lattice, the Fermi energy is shifted towards higher energies. In **Fig. 2** electronic structures of pristine graphene and doped species were also compared. The energy range in this figure is chosen so that the p states of boron atoms can be clearly seen. For the full range DOS plots please see Fig. S1. It is well known that pristine graphene is a semimetal with a zero-band gap [20]. This was also observed here, and it is also shown that doped models behave as conductors. This is very important for the materials' potential electrochemical purposes. After setting the models of boron-doped graphene with various concentrations of B, we move on to exploring their reactivity towards species of interest.

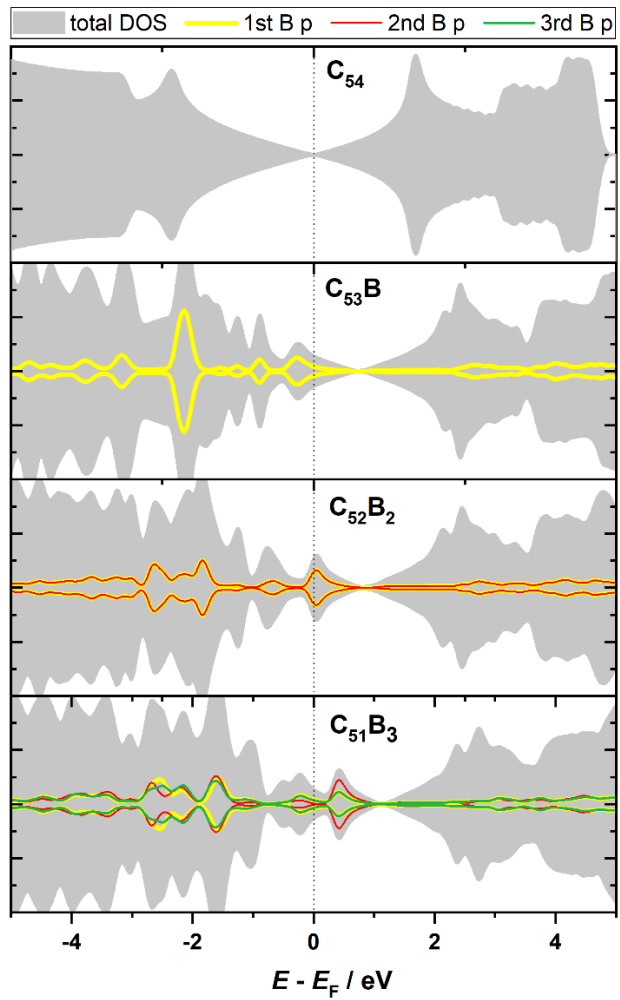

**Fig. 2.** Electronic structures (total DOS) of pristine graphene ($C_{54}$, first row) and investigated boron-doped graphenes $C_{54-n}B_n$ (bottom three rows). The p states of boron dopant atoms are shown separately (in case of $C_{52}B_2$, the p states of the two B atoms overlap). For clarity, all total DOS plots are divided by 5. The Fermi level (dashed, black line) is set to 0.

### 3.2. Reactivity and oxidation

The H, O and OH adsorption trends of $C_{54-n}B_n$ are summarized in **Fig. 3.** When investigating H adsorption (**Fig. 3**, top), the C-top sites were found to be the most favourable ones. For the B-doped surfaces, carbon that is the closest to boron is the one where H prefers to adsorb. This is in agreement with some previous research [12,37]. We find that with the increment of boron percentage in the system, the H adsorption energy decreases, i.e. H is bound more strongly. For pristine graphene the energy of H adsorption is –0.79 eV, while for, $C_{51}B_3$ it's –2.06 eV. The C−H bond length does not change upon B introduction significantly (1.13-1.15 Å).

The situation is somewhat different when looking at the adsorption of O (**Fig. 3**, middle). The preferred site is the bridge site, which in the case of pristine graphene implies the site between two carbon atoms. In the investigated cases of B-doped graphene the preferred bridge site is C−B bridge. The energy of adsorption behaves in the same way as for the H adoption – boron percentage and adsorption energies of O are negatively correlated, the lowest energy being –4.30 eV for O@$C_{51}B_3$.

The adsorption of OH (**Fig. 3**, bottom) gave similar results as in previously done research [12] – OH is binding via creating a C−O bond in the case of pristine graphene, and a B−O bond in the case of $C_{54-n}B_n$. When there are two boron atoms in the model, the adsorption to each one of them gave similar results with the more stable one having the binding energy of –2.21 eV and B−O and O−H bond lengths of 1.52 Å and 0.98 Å, respectively. However, for the $C_{51}B_3$ structure, it is most favourable for the OH group to adsorb on top of the boron atom that is in the middle, obtaining the binding energy of –2.21 eV and the same bond lengths as in $C_{52}B_2$. It is also interesting to note that, unlike the cases where adsorption of H and O was investigated, the most stable model with adsorbed OH is not the $C_{51}B_3$ structure but $C_{52}B_2$. The bond lengths do not vary much when increasing the number of boron atoms. It is also noted that the re-hybridization from $sp^2$ to $sp^3$ occurs in all the cases upon H, O or OH adsorption, meaning that the models lose their initial planarity.

Obviously, all the investigated adsorbates bind more strongly to B-doped graphene than to pristine graphene, with the strength of the interaction rising with B percentage. This means that the introduction of B into the graphene lattice results in enhanced reactivity of the material towards H, O and OH. The increased reactivity towards H is of significance for the possible applications of such materials for hydrogen production and storage. The adsorption energy of atomic hydrogen is used as one of the basic descriptors for hydrogen evolution reaction (HER) activity of the material. In fact, metal-free substitutionally B-doped graphene (1.85 at.% of B) was previously demonstrated to have significantly higher HER activity compared to defective graphene and glassy carbon, with HER onset potential around −0.3 V *vs*. RHE (Reversible Hydrogen Electrode) [38]. This value agrees well with the calculated hydrogen binging energy on B-doped graphene (~ −2.1 eV, **Fig. 3**). Moreover, from the herepresented results, it is clear that the HER active sites are not the B site, but the carbon atoms adjacent to the dopant sites, whose activity is rendered by B incorporation into the graphene lattice.

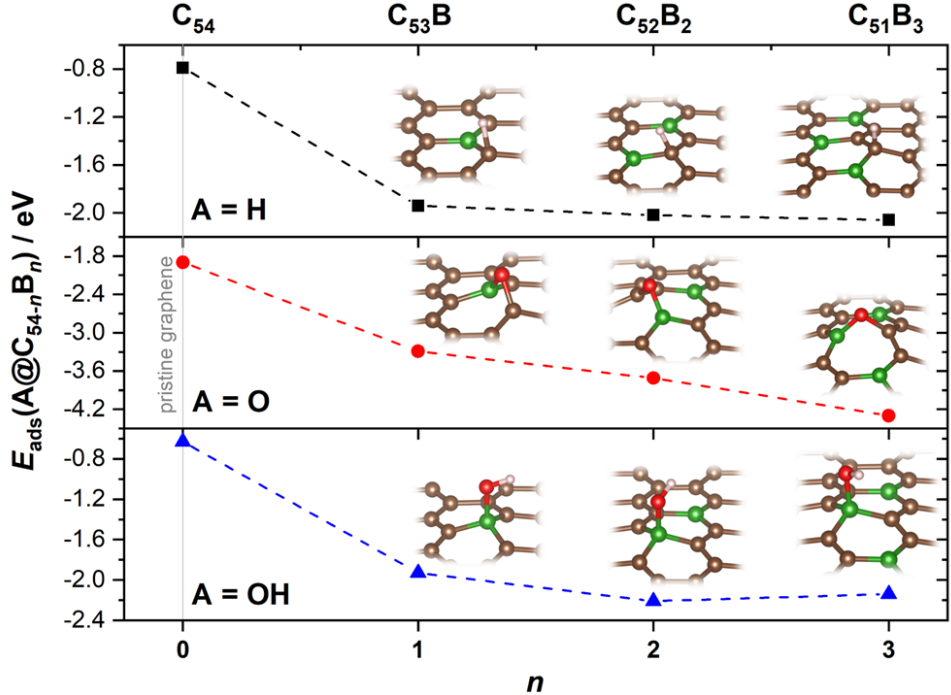

**Fig. 3.** The adsorption trends of H, O and OH on $C_{54-n}B_n$ models: the relationship between the adsorption energies and the number of boron atoms $n$ in the system. For boron-containing models, the optimized adsorption structures are given for each system.

On the other hand, enhanced reactivity of $C_{54-n}B_n$ towards O-containing groups signals that boron doping makes graphene more prone to oxidation, *i.e.* that O-groups will be more stable on the surface containing B, than the pristine one. This is very important from the aspect of metal interaction with such surfaces, which is crucial for applications of materials in metal-ion batteries. It is known that the oxygen-containing groups on graphene can play a very important role in that case, allowing for stronger interaction of chosen metal with the surface [33,35]. However, there is a risk of the O-group interacting "too strongly" with the metal and detaching from the surface irreversibly. Therefore, the stabilization of O-groups on graphene by its doping with B could be a good strategy for better performance of graphene-based materials in metal-ion batteries. Another important question, when it comes to electrochemical applications, is the electrical conductivity of the material. The electronic structures of $C_{54-n}B_n$ upon adsorption of H, O and OH are given in **Fig. 4**. It can be seen that oxidized forms of $C_{54-n}B_n$ (i.e. $C_{54-n}B_n$ with O and OH adsorbed) act as conductors, making them good candidate for further research. The adsorption of two metals of interest, Na and Al, onto bare and oxidized $C_{54-n}B_n$ surfaces will be investigated in the continuation.

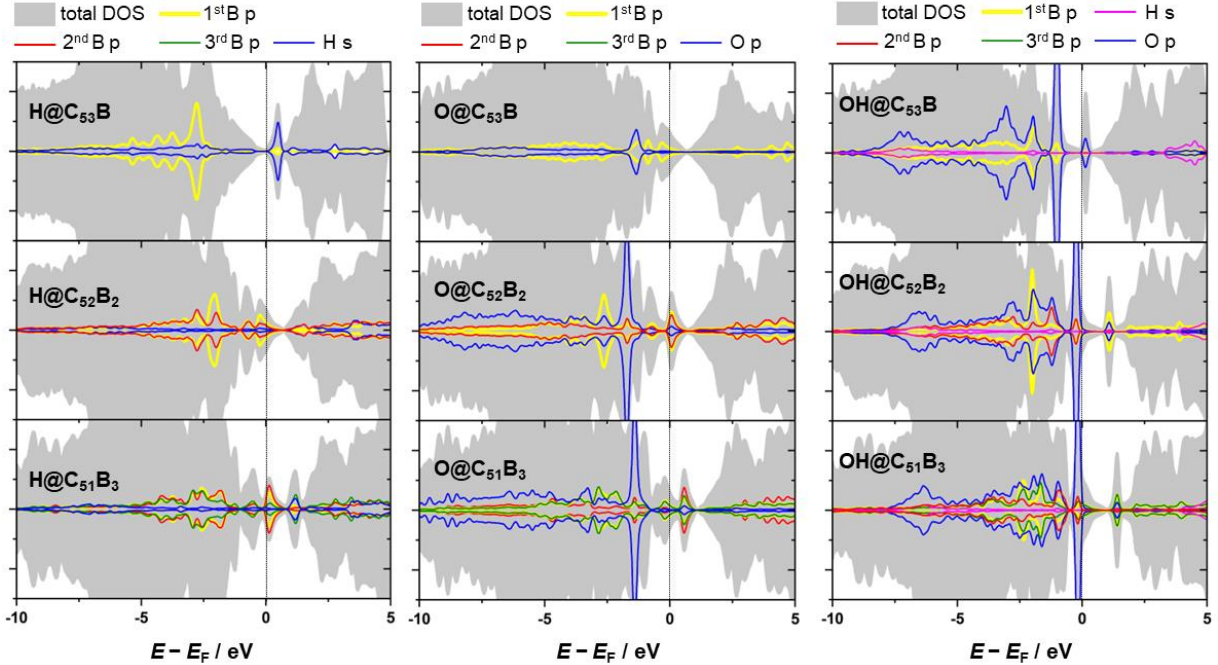

**Fig. 4.** Electronic structures of the optimized H@$C_{54-n}B_n$ (left), O@$C_{54-n}B_n$ (middle) and OH@$C_{54-n}B_n$ systems. Except for the total DOSes and p states of dopant atoms, the states of the adsorbate are also given: p states of O (cases of $O_{ads}$ and $OH_{ads}$) and s states of H (cases of $H_{ads}$ and $OH_{ads}$).

### 3.3. Metal adsorption

Sodium is known to adsorb on pristine graphene at the $C_6$-hollow site, with the adsorption energy of −0.93 eV [15]. Aluminum also adsorbs to hollow site on pristine graphene, with the adsorption energy amounting to −1.09 eV [15]. The reported energies vary depending on the computational approach (especially the chosen dispersion correction) and the size of the supercell. Regarding the adsorption of Na onto an oxidized (nondoped) graphene it is known that Na can directly interact with OH group, resulting in its detachment from the basal plane and NaOH phase separation [39]. We confirm all the mentioned facts regarding pristine graphene here, and additionaly, investigate the adsorption of sodium and aluminum onto $C_{54-n}B_n$, O@$C_{54-n}B_n$ and OH@$C_{54-n}B_n$. The calculated metal adsorption energies on pristine graphene are found to be −0.85 eV and −1.26 eV, for Na and Al, respectively. When it comes to OH@$C_{54}$, we find that both Na and Al adsorption result in metal hydroxyde phase separation. On the other hand, for O@$C_{54}$ there is no detachment of the O-group, and we find that the adsorption energy of Na is −1.54 eV, and of Al −3.84 eV. In case of $C_{54-n}B_n$, the metal can occupy various top, bridge, or hollow adsorption sites. For the oxidized surfaces, the metal generally interacts directly with the oxygen-containing group. Three different outcomes can be identified, and they are represented in **Fig. 5**. Here we have shown only the representatives of the three main groups of adsorption geometries. All optimized structures of Na and Al adsorption on the investigated (oxidized) B-doped graphenes can be found in SI, Figs. S2 and S3.

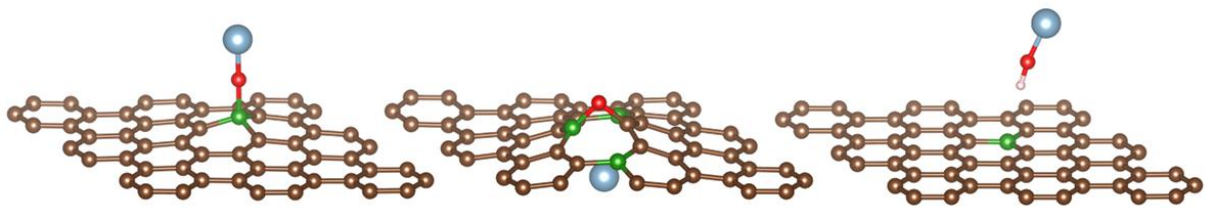

**Fig. 5.** Examples of different outcomes of the metal adsorption on top of the obtained models, for the case of Al: Al adsorbed at O@$C_{53}$B (left), Al at O@$C_{51}B_3$ (middle) and Al at non-oxidized $C_{53}$B (right).

The first outcome is the cases where the metal atom is adsorbed on the same side of the basal plane as O or OH. In these cases, the metal atom interacts directly with the oxygen-containing group. Most of the investigated systems behave like this. In the second outcome, adsorbed Al or Na is on the opposite side of the basal plane from oxygen. This was only found for $C_{51}B_3$ models with an adsorbed oxygen atom. It can be explained by the combination of high dopant concentration and the presence of O-group, which significantly alter the planarity and charge of the surface. Because of this, it is easier for the metal atom to approach the other side of the plane. In both cases, the adsorbed metal occupies a hollow site on the opposite side from $O_{ads}$. This is similar to the case of Na adsorption on corrugated doped graphene [40]. The third outcome represent the cases where the new phase emerges upon the adsorption of the metal. This happens in three cases: Na@OH@pristine graphene, Al@OH@pristine graphene, and when OH is bounded on top of the $C_{53}$B, after the adsorption of Al. Therefore, these new phases represent the formation of metal hydroxyde. These models were not considered further since the formation of a new phase is not favourable for electrochemical purposes. Additionally, charge redistribution is considered through charge difference plots. Again, for brevity, the mentioned plots are given in **Fig. 6** only for the chosen systems, which are representatives of different types of interaction of the metals with the surfaces.

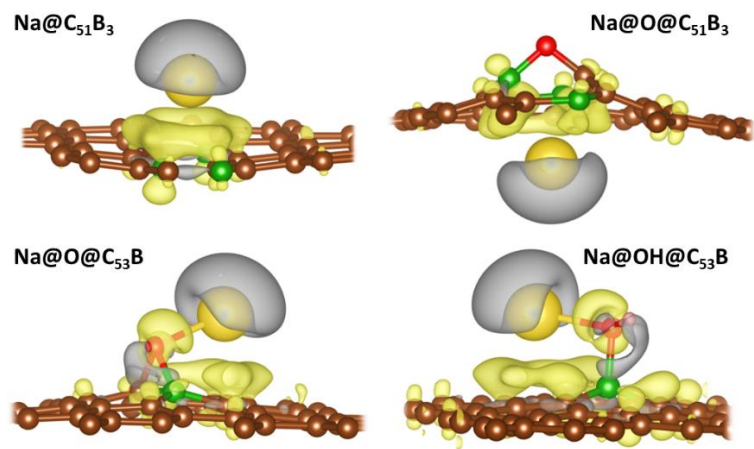

**Fig. 6.** Charge difference plots for the indicated chosen systems of Na adsorption. Grey isosurfaces indicate charge loss, while the yellow ones indicate charge gain. The isosurface values are ±0.002 eV/Å.

It can be clearly seen that in all the cases the metal atoms are the ones that lose their charge and transfer it either completely to the graphene basal plane (cases of non-oxidized surfaces), or partially to oxygen and partially to the basal plane (cases of oxidized surfaces). The only exception when it comes to the oxidized surfaces is the already mentioned M@O@$C_{51}B_3$ (**Fig. 6**, top right), where M adsorbs on the opposite side of the basal plane from O, so that no charge transfer from M to O is found. According to Löwdin charges, Na transfers 0.65-0.75 e to the substrate in all cases, while Al transfers 0.46-0.52 e to the substrate.

In order to avoid forming a metal precipitate, the interaction energy between metal and substrate needs to be higher than the cohesive energies of metals [41]. Thus, one of the main points of interest was comparing the absolute value of calculated adsorption energies with cohesive energies of Na 1.113 eV atom$^{-1}$, and Al 3.39 eV atom$^{-1}$ [42,43], as in **Fig. 7**. It can be seen that in the case of sodium, only one system has an energy lower than the cohesive energy and that's O@$C_{54}$. On the other hand, for aluminium half of the systems, not counting the ones where a new phase emerges, do not fulfil this condition. This is intuitive since Al has higher cohesive energy compared to Na. The systems where the precipitation of Al does not occur are O@$C_{54}$, O@$C_{53}$B, O@$C_{52}B_2$, OH@$C_{52}B_2$, and OH@$C_{51}B_3$.

For electrochemistry purposes "the intermediate" adsorption energy is the one that is preferable, meaning that it is high enough for metals not to precipitate but the adsorption is not so strong that it becomes irreversible.

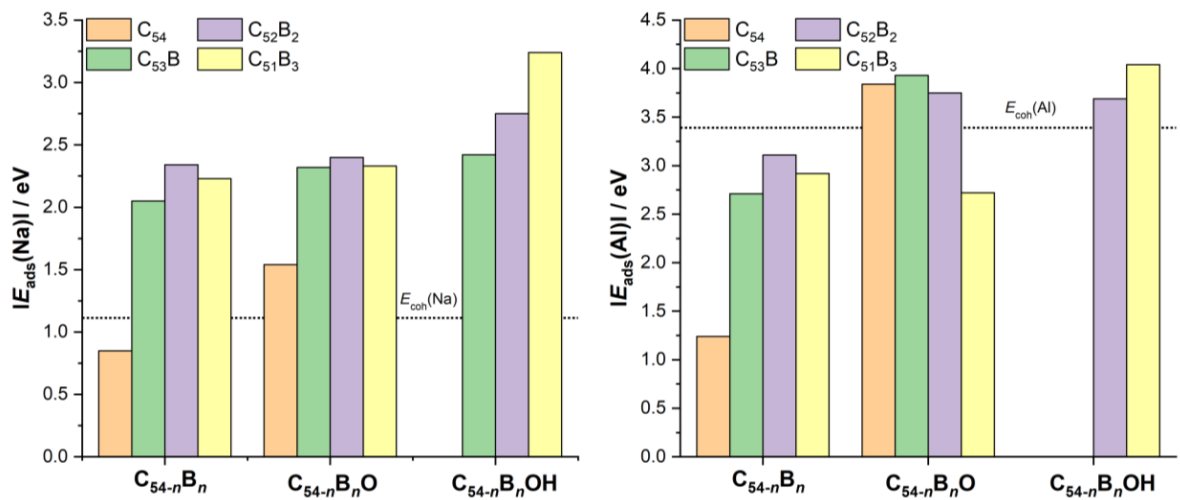

**Fig. 7.** Adsorption energies of Na (left) and Al (right) on $C_{54-n}B_n$ models, compared to their cohesive energies ($E_{coh}$), given by dotted lines. The missing columns (Na@$C_{54}$OH, Al@$C_{54}$OH and Al@$C_{53}$BOH) indicate metal-hydroxyde phase separation.

For the substrates without pre-adsorbed O or OH, the absolute value of the adsorption energies of the sodium and aluminium increases with the increment of boron atoms in the structure up to a point where there are two boron atoms, and then it decreases again, making the $C_{52}B_2$ model the one with the most exothermic adsorption energy for both metals. The same trend is observed for Na when O is pre-adsorbed on the surface. However, this is not the case for Al where the highest absolute value of the adsorption energy is for the O@$C_{52}B_2$ structure. When OH is bonded to the surface, for both metals the absolute value of the adsorption energy increases with the percentage of boron.

Another important point is determining whether the systems are conducting. They need to be conducting to be used electrode materials. This is why the analysis of the electronic structure, in particular, the density of states (DOS) was done for all the investigated systems. **Fig. 8** shows them for the case of sodium (for brevity). Out of the considered models that fulfil previous conditions, almost all behave as conductors except $C_{53}B$ and O@$C_{53}B$ which upon adsorbing sodium have very low DOS around the Fermi energy (**Fig. 8**).

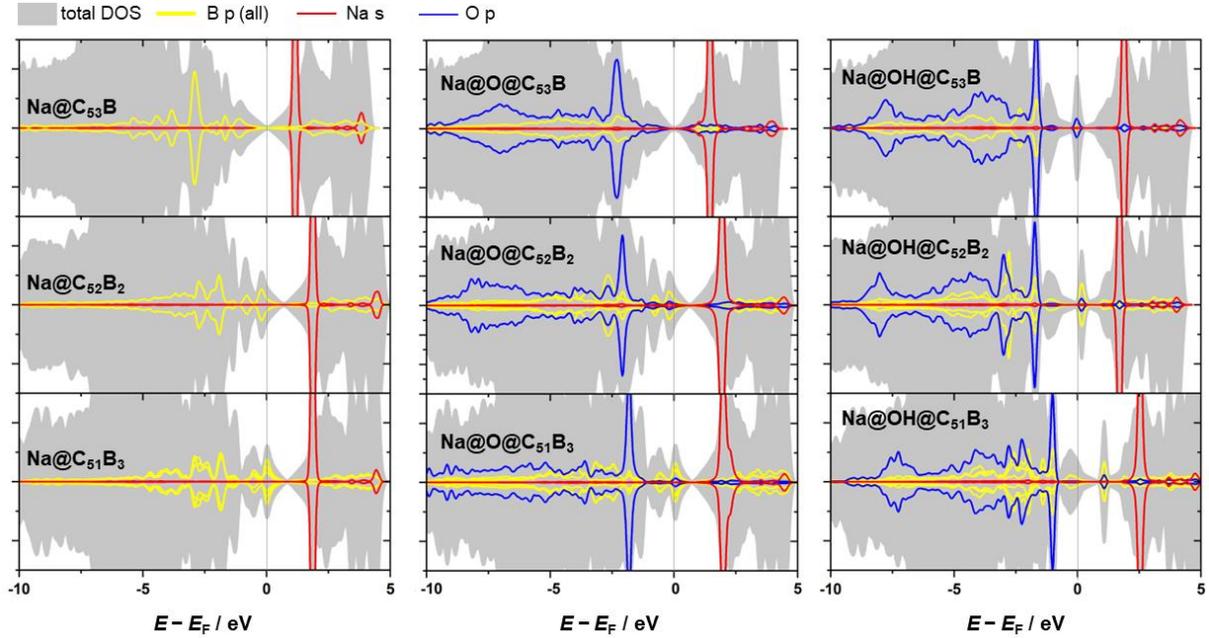

**Fig. 8.** Electronic structures of sodium adsorption onto non-oxidized and oxidized $C_{54-n}B_n$ models. Except for total DOS, the s states of sodium are given, as well as the p states of boron dopant atoms, and p states of O (when present, cases of O@ $C_{54-n}B_n$ and OH@$C_{54-n}B_n$).

Next, we investigated how the presence of boron dopant atoms in different concentrations, as well as the presence of O/OH groups on the surface, affect the adsorption capacity for Na and Al (**Fig. 9**). We compare the calculated differential adsorption energies of each added Na to its cohesive energy and find that up to 4 Na can be adsorbed onto $C_{51}B_3$, O@$C_{53}B$ and O@$C_{52}B_2$. For the oxidized forms of $C_{51}B_3$, the absolute value of $E_{ads,diff}$ of the 4$^{th}$ Na is lower than Na's cohesive energy, which indicates that metal phase precipitation would occur. On the other hand, for OH@$C_{53}B$, Na-O phase separation occurs upon the addition of the 2$^{nd}$ Na, while for $C_{53}B$ and OH@$C_{52}B_2$, it happens upon adding the 3$^{rd}$ Na, and for $C_{52}B_2$, after the addtion of the 4$^{th}$ Na. The optimized structures of 2, 3 and 4 Na adsorption onto the models for which no phase separation occurs and the adsorption energies overcome the cohesive energy of Na are shown in **Fig. 10**. From this figure, it is obvious that, in the case of non-oxidized boron-doped graphene, Na prefers adsorbing on B-containing hollow sites. On the other hand, in the case of the oxidized surfaces, it mostly interacts with the O-group, which can bind up to 3 Na.

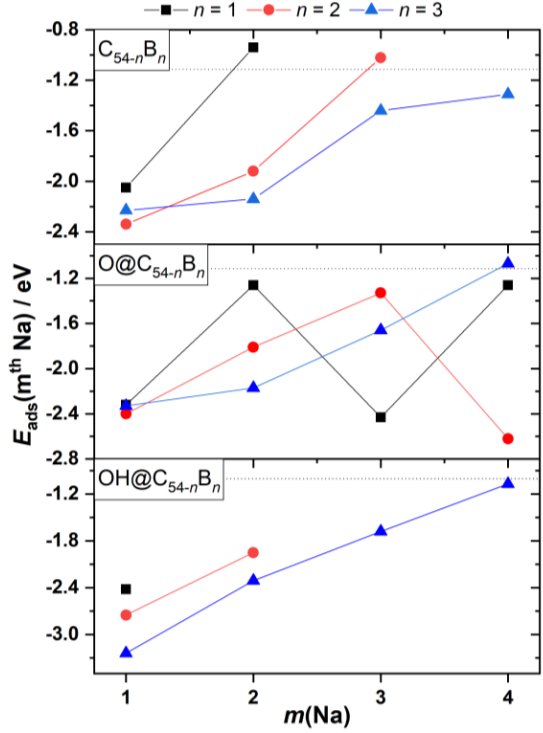

**Fig. 9.** Differential adsorption energies of $m$ Na atoms ($m$ goes from 1 to 4) onto the investigated substrates. Lacking data points indicate M-O phase separation (*e.g.* in the case of 2Na@OH@$C_{53}$B). For comparison, the cohesive energy of Na is given as dotted, grey horizontal lines.

When it comes to Al adsorption (**Table 2**), upon the addition of the 2$^{nd}$ Al, most cases result in adsorption energies lower than the Al cohesive energy (comparing absolute values), while for others Al-O phase separation occurs. The only case with a stable adsorption of 2Al is 2Al@O@$C_{52}B_2$, which also results in Al-O phase separation upon the addition of the third Al atom.

**Table 2.** Differential adsorption energies of 1-2 Al atoms onto the investigated substrates. Lacking values indicate M-O phase separation. Bolded values indicate that the adsorption energy overcomes Al's cohesive energy.

| | $E_{ads}(m^{th}\,Al@subs)$ / eV | | | | | | | | |
|---|---|---|---|---|---|---|---|---|---|
| $m$(Al) | $C_{53}$B | $C_{52}B_2$ | $C_{51}B_3$ | O@$C_{53}$B | O@$C_{52}B_2$ | O@$C_{51}B_3$ | OH@$C_{53}$B | OH@$C_{52}B_2$ | OH@$C_{51}B_3$ |
| 1 | −2.71 | −3.11 | −2.92 | **−3.93** | **−3.75** | −2.72 | **−3.80** | **−3.69** | **−4.05** |
| 2 | - | - | - | - | **−3.70** | - | - | - | −2.81 |

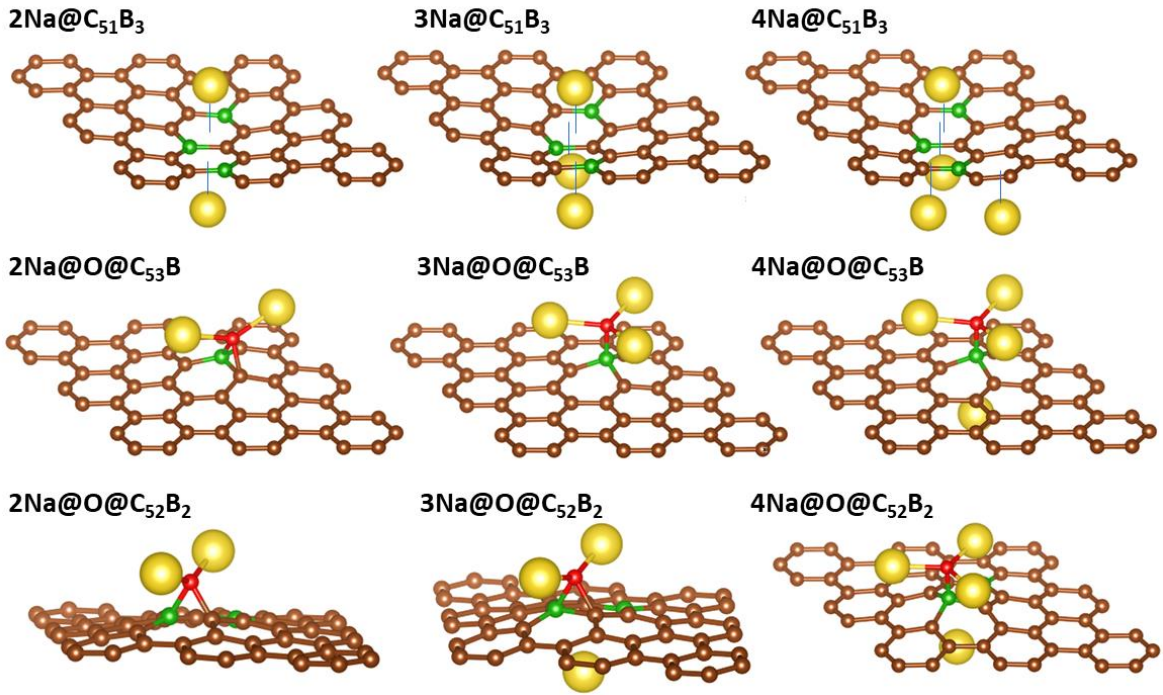

**Fig. 10.** Optimized structures of 2-4 Na adsorption onto $C_{51}B_3$ (top row), $O@C_{53}B$ (middle row), and $O@C_{52}B_2$ (bottom row).

## 4. Conclusion

DFT calculations were utilized to investigate the influence of boron dopant concentration and surface oxidation on the potential for Na and Al storage in graphene. The obtained results demonstrated that the introduction of boron into graphene has a significant effect on its reactivity towards oxygen-containing species, rendering the material more susceptible to oxidation. Consequently, when modeling such materials, it is crucial to consider the oxidation of the B centers. In terms of electrochemical applications, the oxidation of B-doped graphene modifies its electronic structure but does not induce a bandgap. As a result, controllable doping of graphene with boron, followed by oxidation, appears to be a promising approach for enhancing electrochemical performance with respect to Na and Al storage. Specifically, this strategy optimizes the strength of the interaction between Na and Al and the graphene surface and its storage capacity, thereby opening up possibilities for applications in metal-ion batteries and energy storage in general. In both cases, the adsorbate atoms tend to bind to or near B atoms, which are additionally functionalized by surface groups such as O or OH.

**Acknowledgement**

This work was supported by the Serbian Ministry of Science, Technological Development, and Innovations (contract number: 451-03-47/2023-01/200146). I.A.P. is also indebted to the Research Fund of the Serbian Academy of Sciences and Arts, project F-190, for supporting this study. The computations and data handling were enabled by resources provided by the Swedish National Infrastructure for Computing (SNIC) at the National Supercomputer Centre (NSC) at Linköping University, partially funded by the Swedish Research Council through grant agreement No. 2018-05973.

**References**

1. Novoselov, K.S.; Geim, A.K.; Morozov, S. V.; Jiang, D.; Katsnelson, M.I.; Grigorieva, I. V.; Dubonos, S. V.; Firsov, A.A. Two-Dimensional Gas of Massless Dirac Fermions in Graphene. *Nature* **2005**, *438*, doi:10.1038/nature04233.

2. Liu, J.; Liang, T.; Tu, R.; Lai, W.; Liu, Y. Redistribution of π and σ Electrons in Boron-Doped Graphene from DFT Investigation. *Appl. Surf. Sci.* **2019**, *481*, 344–352, doi:10.1016/j.apsusc.2019.03.109.

3. Brownson, D.A.C.; Kampouris, D.K.; Banks, C.E. An Overview of Graphene in Energy Production and Storage Applications. *J. Power Sources* 2011, *196*.

4. Shanmugam, S.; Nachimuthu, S.; Subramaniam, V. DFT Study of Adsorption of Ions on Doped and Defective Graphene. *Mater. Today Commun.* **2020**, *22*, doi:10.1016/j.mtcomm.2019.100714.

5. Ali, S.; Lone, B. Adsorption of Cytosine on Si and Ge Doped Graphene: A DFT Study. *Mater. Today Proc.* **2023**, *80*, 774–781, doi:10.1016/j.matpr.2022.11.085.

6. Tiwari, S.K.; Sahoo, S.; Wang, N.; Huczko, A. Graphene Research and Their Outputs: Status and Prospect. *J. Sci. Adv. Mater. Devices* 2020, *5*.

7. Soldano, C.; Mahmood, A.; Dujardin, E. Production, Properties and Potential of Graphene. *Carbon N. Y.* 2010, *48*.

8. Agnoli, S.; Favaro, M. Doping Graphene with Boron: A Review of Synthesis Methods, Physicochemical Characterization, and Emerging Applications. *J. Mater. Chem. A* 2016, *4*.

9. Liu, L.; Qing, M.; Wang, Y.; Chen, S. Defects in Graphene: Generation, Healing, and Their Effects on the Properties of Graphene: A Review. *J. Mater. Sci. Technol.* **2015**, *31*, doi:10.1016/j.jmst.2014.11.019.

10. Yang, G.; Li, L.; Lee, W.B.; Ng, M.C. Structure of Graphene and Its Disorders: A Review. *Sci. Technol. Adv. Mater.* 2018, *19*.

11. Riyaz, M.; Garg, S.; Kaur, N.; Goel, N. Boron Doped Graphene as Anode Material for Mg Ion Battery: A DFT Study. *Comput. Theor. Chem.* **2022**, *1214*, 113757,

doi:10.1016/j.comptc.2022.113757.

12. Dobrota, A.S.; Pašti, I.A.; Mentus, S. V.; Skorodumova, N. V. A DFT Study of the Interplay between Dopants and Oxygen Functional Groups over the Graphene Basal Plane - Implications in Energy-Related Applications. *Phys. Chem. Chem. Phys.* **2017**, *19*, doi:10.1039/c7cp00344g.

13. Zhang, Q.; Zhou, Y.; Yu, Y.; Chen, B.-Y.; Hong, J. Exploring Catalytic Performance of Boron-Doped Graphene Electrode for Electrochemical Degradation of Acetaminophen. *Appl. Surf. Sci.* **2020**, *508*, 145111, doi:10.1016/j.apsusc.2019.145111.

14. Etacheri, V.; Marom, R.; Elazari, R.; Salitra, G.; Aurbach, D. Challenges in the Development of Advanced Li-Ion Batteries: A Review. *Energy Environ. Sci.* 2011, *4*.

15. Pašti, I.A.; Jovanović, A.; Dobrota, A.S.; Mentus, S. V.; Johansson, B.; Skorodumova, N. V. Atomic Adsorption on Pristine Graphene along the Periodic Table of Elements – From PBE to Non-Local Functionals. *Appl. Surf. Sci.* **2018**, *436*, doi:10.1016/j.apsusc.2017.12.046.

16. Lin, M.C.; Gong, M.; Lu, B.; Wu, Y.; Wang, D.Y.; Guan, M.; Angell, M.; Chen, C.; Yang, J.; Hwang, B.J.; et al. An Ultrafast Rechargeable Aluminium-Ion Battery. *Nature* **2015**, *520*, doi:10.1038/nature14340.

17. Zhou, W.; Jia, J.; Lu, J.; Yang, L.; Hou, D.; Li, G.; Chen, S. Recent Developments of Carbon-Based Electrocatalysts for Hydrogen Evolution Reaction. *Nano Energy* **2016**, *28*, 29–43, doi:10.1016/j.nanoen.2016.08.027.

18. Panchakarla, L.S.; Subrahmanyam, K.S.; Saha, S.K.; Govindaraj, A.; Krishnamurthy, H.R.; Waghmare, U. V.; Rao, C.N.R. Synthesis, Structure, and Properties of Boron- and Nitrogen-Doped Graphene. *Adv. Mater.* **2009**, *21*, doi:10.1002/adma.200901285.

19. Yu, X.; Han, P.; Wei, Z.; Huang, L.; Gu, Z.; Peng, S.; Ma, J.; Zheng, G. Boron-Doped Graphene for Electrocatalytic N2 Reduction. *Joule* **2018**, *2*, 1610–1622, doi:10.1016/j.joule.2018.06.007.

20. Rao, C.N.R.; Gopalakrishnan, K.; Govindaraj, A. Synthesis, Properties and Applications of Graphene Doped with Boron, Nitrogen and Other Elements. *Nano Today* 2014, *9*.

21. Liu, Y.; Artyukhov, V.I.; Liu, M.; Harutyunyan, A.R.; Yakobson, B.I. Feasibility of Lithium Storage on Graphene and Its Derivatives. *J. Phys. Chem. Lett.* **2013**, *4*, doi:10.1021/jz400491b.

22. Hao, Q.; Xia, X.; Lei, W.; Wang, W.; Qiu, J. Facile Synthesis of Sandwich-like Polyaniline/Boron-Doped Graphene Nano Hybrid for Supercapacitors. *Carbon N. Y.* **2015**, *81*, doi:10.1016/j.carbon.2014.09.090.

23. Zhang, L.; Zhang, Z.Y.; Liang, R.P.; Li, Y.H.; Qiu, J.D. Boron-Doped Graphene Quantum Dots for Selective Glucose Sensing Based on the "Abnormal" Aggregation-Induced Photoluminescence Enhancement. *Anal. Chem.* **2014**, *86*, doi:10.1021/ac500289c.

24. Choudhuri, I.; Patra, N.; Mahata, A.; Ahuja, R.; Pathak, B. B-N@Graphene: Highly Sensitive and Selective Gas Sensor. *J. Phys. Chem. C* **2015**, *119*, doi:10.1021/acs.jpcc.5b07359.

25. Fan, Z.; Li, Y.; Li, X.; Fan, L.; Zhou, S.; Fang, D.; Yang, S. Surrounding Media Sensitive Photoluminescence of Boron-Doped Graphene Quantum Dots for Highly Fluorescent Dyed

Crystals, Chemical Sensing and Bioimaging. *Carbon N. Y.* **2014**, *70*, doi:10.1016/j.carbon.2013.12.085.

26. Ling, C.; Mizuno, F. Boron-Doped Graphene as a Promising Anode for Na-Ion Batteries. *Phys. Chem. Chem. Phys.* **2014**, *16*, doi:10.1039/c4cp01045k.

27. Du, Y.; Zhang, B.; Kang, R.; Zhou, W.; Zhang, W.; Jin, H.; Wan, J.; Zhang, J.; Chen, G. Boron-Doping-Induced Defect Engineering Enables High Performance of a Graphene Cathode for Aluminum Batteries. *Inorg. Chem. Front.* **2022**, *9*, doi:10.1039/d1qi01474a.

28. Diklić, N.P.; Dobrota, A.S.; Pašti, I.A.; Mentus, S. V.; Johansson, B.; Skorodumova, N. V. Sodium Storage via Single Epoxy Group on Graphene – The Role of Surface Doping. *Electrochim. Acta* **2019**, *297*, doi:10.1016/j.electacta.2018.11.108.

29. Giannozzi, P.; Baroni, S.; Bonini, N.; Calandra, M.; Car, R.; Cavazzoni, C.; Ceresoli, D.; Chiarotti, G.L.; Cococcioni, M.; Dabo, I.; et al. QUANTUM ESPRESSO: A Modular and Open-Source Software Project for Quantum Simulations of Materials. *J. Phys. Condens. Matter* **2009**, *21*, doi:10.1088/0953-8984/21/39/395502.

30. Giannozzi, P.; Andreussi, O.; Brumme, T.; Bunau, O.; Buongiorno Nardelli, M.; Calandra, M.; Car, R.; Cavazzoni, C.; Ceresoli, D.; Cococcioni, M.; et al. Advanced Capabilities for Materials Modelling with Quantum ESPRESSO. *J. Phys. Condens. Matter* **2017**, *29*, doi:10.1088/1361-648X/aa8f79.

31. S. Grimme Semiempirical GGA-Type Density Functional Constructed with a Long-Range Dispersion Correction. *J. Comput. Chem.* **2006**, *27*, doi:10.1002/jcc.20495.

32. Monkhorst, H.J.; Pack, J.D. Special Points for Brillouin-Zone Integrations. *Phys. Rev. B* **1976**, *13*, doi:10.1103/PhysRevB.13.5188.

33. Momma, K.; Izumi, F. VESTA 3 for Three-Dimensional Visualization of Crystal, Volumetric and Morphology Data. *J. Appl. Crystallogr.* **2011**, *44*, doi:10.1107/S0021889811038970.

34. Dobrota, A.S.; Pašti, I.A.; Skorodumova, N. V. Oxidized Graphene as an Electrode Material for Rechargeable Metal-Ion Batteries - a DFT Point of View. *Electrochim. Acta* **2015**, *176*, doi:10.1016/j.electacta.2015.07.125.

35. Dobrota, A.S.; Gutić, S.; Kalijadis, A.; Baljozović, M.; Mentus, S. V.; Skorodumova, N. V.; Pašti, I.A. Stabilization of Alkali Metal Ions Interaction with OH-Functionalized Graphene: Via Clustering of OH Groups-Implications in Charge Storage Applications. *RSC Adv.* **2016**, *6*, doi:10.1039/c6ra13509a.

36. Pašti, I.A.; Jovanović, A.; Dobrota, A.S.; Mentus, S. V.; Johansson, B.; Skorodumova, N. V. Atomic Adsorption on Graphene with a Single Vacancy: Systematic DFT Study through the Periodic Table of Elements. *Phys. Chem. Chem. Phys.* **2018**, *20*, doi:10.1039/c7cp07542a.

37. Miwa, R.H.; Martins, T.B.; Fazzio, A. Hydrogen Adsorption on Boron Doped Graphene: An Ab Initio Study. *Nanotechnology* **2008**, *19*, doi:10.1088/0957-4484/19/15/155708.

38. Sathe, B.R.; Zou, X.; Asefa, T. Metal-Free B-Doped Graphene with Efficient Electrocatalytic Activity for Hydrogen Evolution Reaction. *Catal. Sci. Technol.* **2014**, *4*, doi:10.1039/c4cy00075g.

39. Dobrota, A.S.; Pašti, I.A.; Mentus, S. V.; Johansson, B.; Skorodumova, N. V. Functionalized Graphene for Sodium Battery Applications: The DFT Insights. *Electrochim. Acta* **2017**, *250*, doi:10.1016/j.electacta.2017.07.186.

40. Dobrota, A.S.; Pašti, I.A.; Mentus, S. V.; Johansson, B.; Skorodumova, N. V. Altering the Reactivity of Pristine, N- and P-Doped Graphene by Strain Engineering: A DFT View on Energy Related Aspects. *Appl. Surf. Sci.* **2020**, *514*, doi:10.1016/j.apsusc.2020.145937.

41. Oh, J.S.; Kim, K.N.; Yeom, G.Y. Graphene Doping Methods and Device Applications. *J. Nanosci. Nanotechnol.* 2014, *14*.

42. Kittel, C. *Introduction to Solid State Physics, 8th Edition*; Wiley & Sons, 2004;

43. Kaxiras, E. *Atomic and Electronic Structure of Solids*; Cambridge University Press, 2010;

**SUPPLEMENTARY INFORMATION**

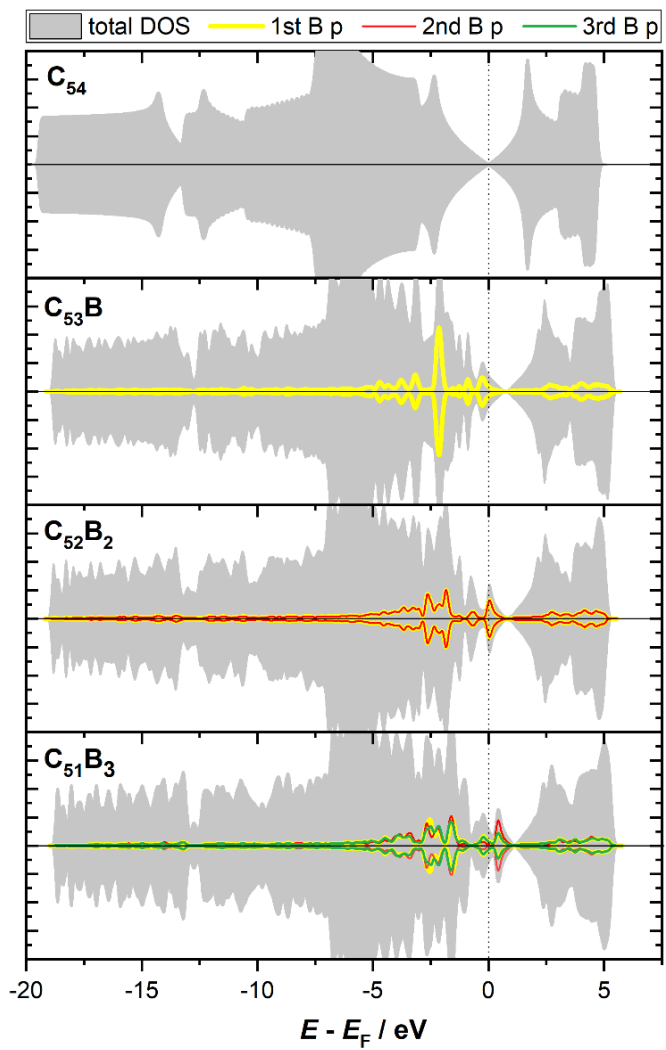

**Fig. S1.** Electronic structures (total DOS) of pristine graphene ($C_{54}$, first row) and investigated boron-doped graphenes $C_{54-n}B_n$ (bottom three rows), in full energy range. The p states of boron dopant atoms are shown separately (in case of $C_{52}B_2$, the p states of two B atoms overlap). For clarity, all total DOS plots are divided by 5. The Fermi level (dashed, black line) is set to 0.

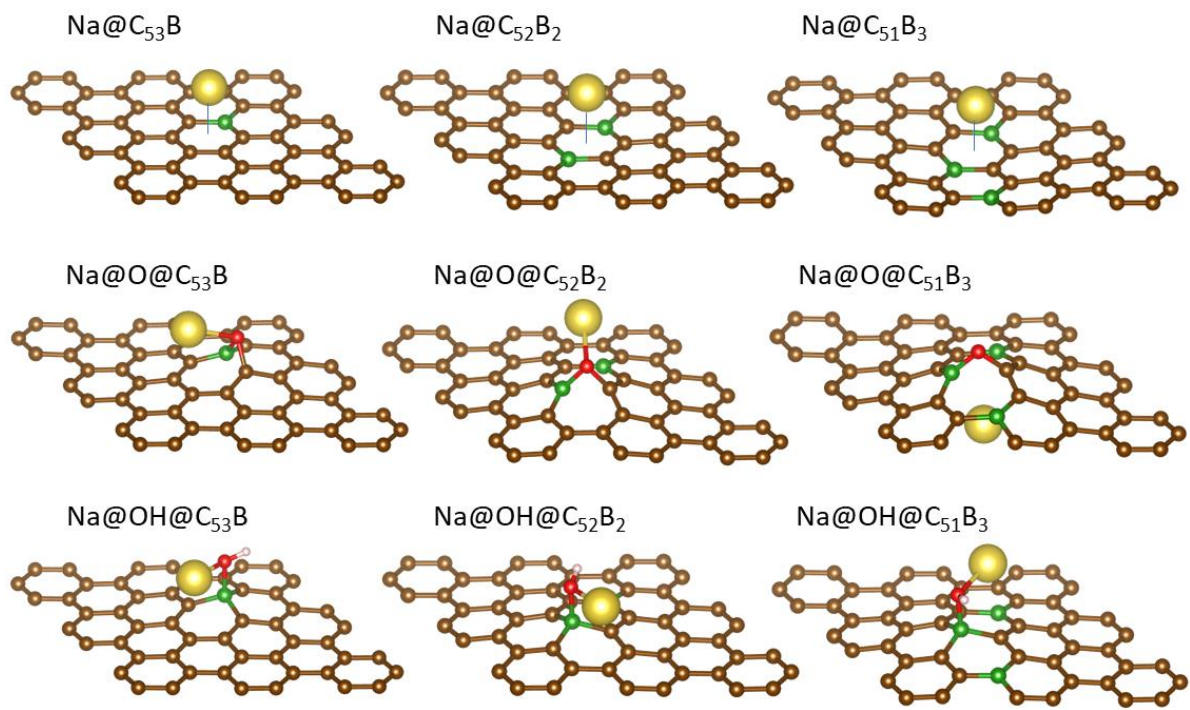

**Fig. S2.** Optimized structures of Na adsorption onto (oxidized) $C_{54-n}B_n$ systems (first row: bare $C_{54-n}B_n$, middle row $C_{54-n}B_n$ oxidized by $O_{ads}$, bottom row $C_{54-n}B_n$ oxidized by $OH_{ads}$). Graphical representations was made using VESTA.

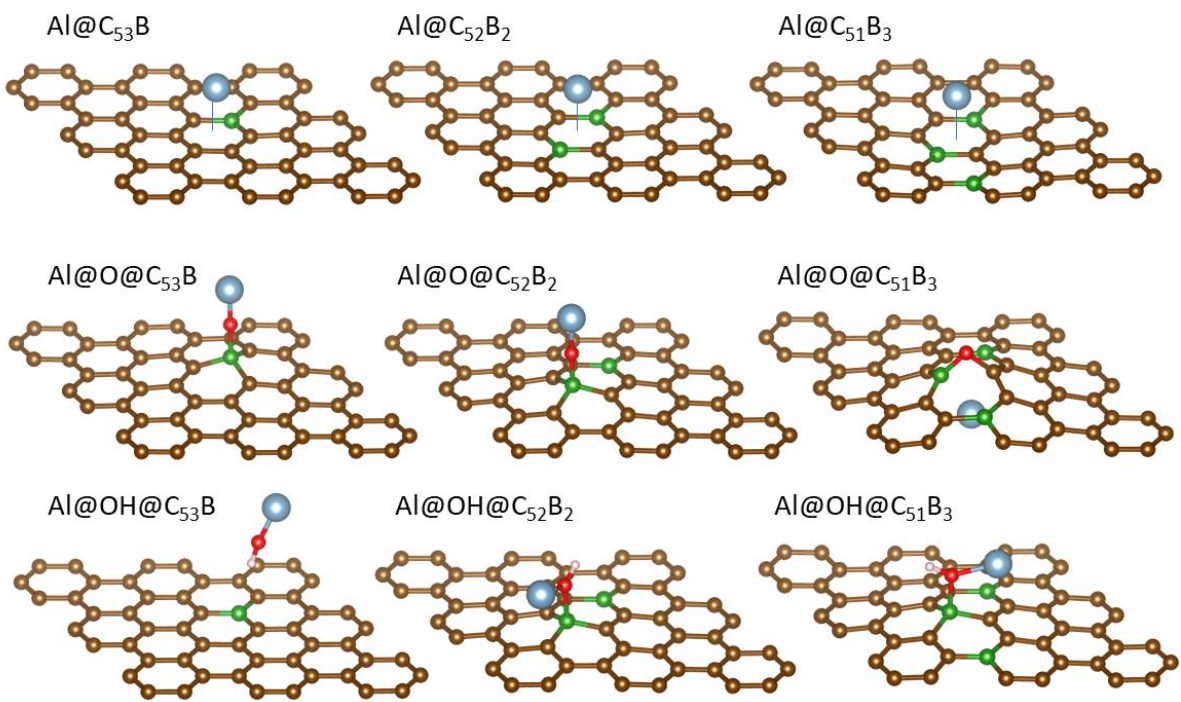

**Fig. S3.** Optimized structures of Al adsorption onto (oxidized) $C_{54-n}B_n$ systems (first row: bare $C_{54-n}B_n$, middle row $C_{54-n}B_n$ oxidized by $O_{ads}$, bottom row $C_{54-n}B_n$ oxidized by $OH_{ads}$). Graphical representations were made using VESTA.